\documentclass[onecolumn,amsmath,amssymb,a4paper,preprint,superscriptaddress]{revtex4}
\usepackage[dvipdfm]{graphicx}
\usepackage{natbib}
\usepackage{multirow}
\usepackage{amsmath}
\usepackage{url}

\begin{document}
\title{ A direct solution to the phonon Boltzmann equation}

\author{Laurent Chaput}
\email{laurent.chaput@ijl.nancy-universite.fr}
\affiliation{Institut Jean Lamour, UMR CNRS 7198, Nancy Universit\'{e}, Bd. des Aiguillettes, BP 23, 54506 Vandoeuvre Les Nancy Cedex, France}

\begin{abstract}
The frequency dependent phonon Boltzmann equation is transformed to an integral equation over the irreducible part of the Brillouin zone. Simultaneous diagonalization of the collision kernel of that equation and a symmetry crystal class operator allow to obtain a spectral representation of the lattice thermal conductivity valid at finite frequency. The method is applied to C, Si and Mg$_2$Si to obtain the static and dynamical thermal conductivities.
\end{abstract}

\maketitle

\begin{figure}
\begin{center}
\includegraphics[scale=0.35]{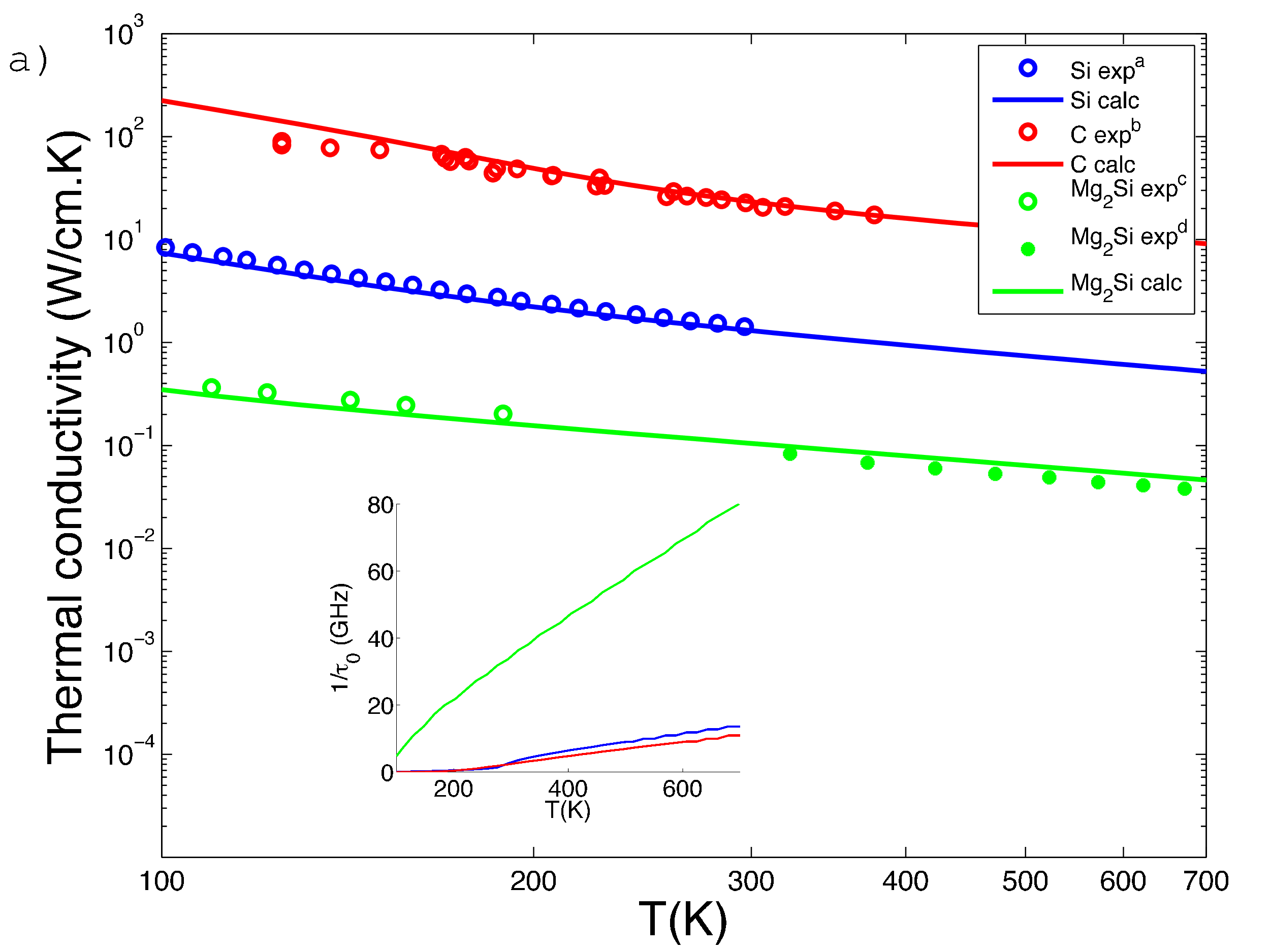}
\includegraphics[scale=0.35]{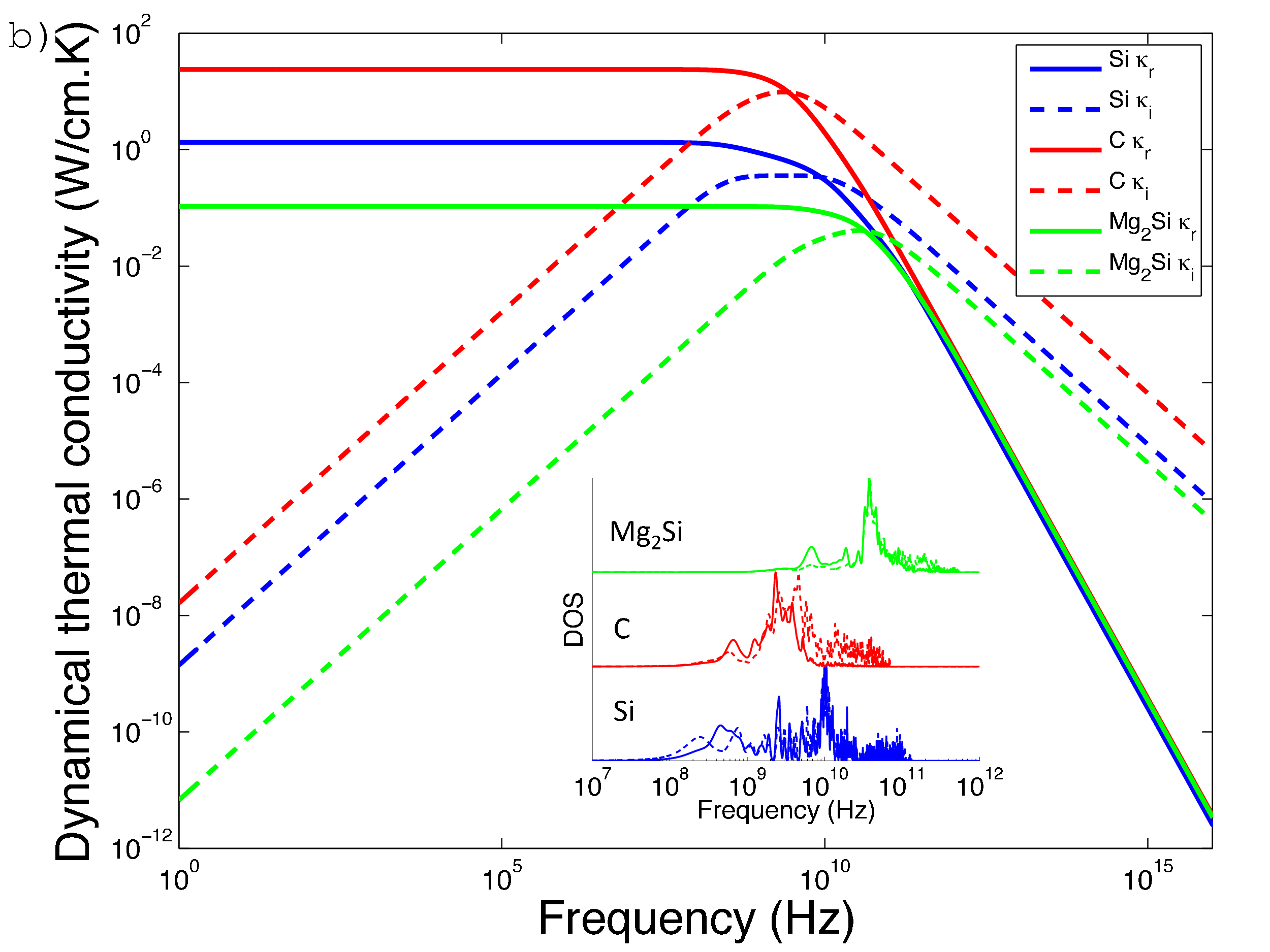}
\caption{Lattice thermal conductivities for diamond (red), Si (blue) and Mg$_2$Si (green). The panel a) shows the calculated (full lines) and experimental (circles) static thermal conductivity as a function of temperature. Panel b) shows the real (full lines) and imaginary (dashed lines) parts of the dynamical lattice thermal conductivity. The inset in  panel a) shows the inverse relaxation time $\tau_0$ as a function of temperature. The inset in panel b) shows spectral densities. The full line represent $\rho_{\alpha \beta}$ while for the dashed line is just $\propto \sum_{r} \delta(\omega -\omega_r)$.  The experimental data are taken from  $^a$Ref\cite{inyushkini} $^b$Ref\cite{wei} $^c$Ref\cite{martin} $^d$Ref\cite{akasaka}. \label{fig:1}}
\end{center}
\end{figure}

The study of lattice heat transport in a crystal compound requires the knowledge of the phonon excitations as well as a model of transport which, in bulk systems, is conveniently taken to be the Boltzmann equation. Harmonic phonon spectrum are routinely obtained from \emph{ab-initio} calculations  based on the density functional theory, even for complicated compounds\cite{euchner12}.  However to access transport properties like the thermal conductivity such calculations are not sufficient since it is necessary to describe the scattering of harmonic phonons by others phonons, impurities and crystal boundaries. Among those three scattering processes the scattering by the others phonons is usually the more demanding since it originate in the anharmonic part of the total energy and therefore at least the third variation of the energy with respect to atomic displacements is needed. Such calculations have however be shown to be feasible,  either from density perturbation theory\cite{deinzer}, or using finite displacements\cite{chaput11}.\\
Once these scattering processes are calculated, and the related collision matrices constructed, the transport Boltzmann equation still remains to be solved.
Among the few published results for the \emph{ab-initio} calculations of the thermal conductivity, all solve this equation iteratively and consider the stationary case\cite{ward09}.  It should be noticed that even in the simple case of silicon this scheme requires already several tens of iterations to get the desired accuracy. In addition there is no proof for the convergence of the iteration procedure. However in applications there are many cases of interest where materials with a low thermal conductivity are needed and therefore where such an approach may be inadequate. It is for example the case in thermoelectricity where the lattice thermal conductivity should be as small as possible to increase the figure of merit $ZT$. In this context the study of time/frequency dependent lattice thermal conductivity is also important and is the subject of intense research\cite{koh07}. For a material to be a good thermoelectric it should have good electrical properties such that conductivity and thermopower but should also be a poor thermal conductor. These two requirements have been proved to be difficult to be achieved together in bulk materials. One way to circumvent the difficulty is to nano-structurate the materials\cite{hbrow}. An other maybe to use finite frequency properties . It is known\cite{chester} that at a certain frequency the thermal conductivity start dropping rapidly what will increase the thermoelectric efficiency.  The dropping frequency will be calculated here for the first time using \emph{ab-initio} calculations.\\
In this letter I present a direct non iterative solution to the Boltzmann equation applicable to the stationary and non stationary regime. This allow to calculate the static and dynamical thermal conductivities and the accuracy of the method allows to study materials which are poor thermal conductors. A single parameter control the accuracy, mainly the number of points used to sample the first Brillouin zone. The paper is organized as follow. The Boltzmann equation is first reduced to an integral equation over the irreducible part of the Brillouin zone. Then the thermal conductivity is expressed in term of the collision operator defined during the reduction process. This operator is symmetric and, due to the reduction over the irreducible part of the Brillouin zone, is small enough to be diagonalized numerically. Consequently a spectral representation is obtained for the thermal conductivity, valid at zero and finite frequency.  Finally the method is applied to compounds having from very large to very low thermal conductivity. In each case the agreement with experiment is excellent.\\
The phonon Boltzmann equation is an integral equation over the first Brillouin zone which in its linearized version takes the form \cite{ziman}
\begin{align}
\frac{\partial n_{qp}^{(1)}}{\partial t}+\frac{\partial n_{qp}^{(0)}}{\partial T}\frac{\partial T}{\partial \vec{r}} \cdot \vec{v}_{qp}=  C(qp;n_{qp}^{(1)})+ \frac{1}{2}D(qp;n_{qp}^{(1)}) \label{eq:1}
\end{align}
The scattering of phonon appears in the right hand side of the equation through the term $C(qp)$ for the collision processes, and with $D(qp)$ for the decay processes.
In the above equations $n_{qp}$ is the occupation function for a phonon of wave vector $q$ in branch $p$. $\vec{v}_{qp}$ is the velocity and $T=T(\vec{r},t)$ the temperature. $n_{qp}^{(0)}$ is the occupation function at equilibrium, and $n_{qp}^{(1)}$ is the first order deviation from equilibrium, $n_{qp}\approx n_{qp}^{(0)}+n_{qp}^{(1)}$.
It is possible to rearrange the scattering integral of ref \cite{ziman} in order to make its relation to the lifetime of phonons calculated in \cite{chaput11} more explicit\cite{srivastava},
\begin{align}
C(qp;n_{qp}^{(1)})+ \frac{1}{2}D(qp;n_{qp}^{(1)})=-\sum_{q'p'} \Omega'_{qp,q'p'}n_{q'p'}^{(1)}\frac{ \sinh \Big( \frac{ \hbar \omega_{q'p'}}{2 k_B T}\Big)}{\sinh \Big( \frac{ \hbar \omega_{qp}}{2 k_B T}\Big)}
\end{align}
with 
\begin{align*}
 \Omega'_{qp,q'p'}&=- \frac{\pi}{\hbar^2}\sum_{q_bp_b} \Big| F_{q -q' q_b}^{p p' p_b} \Big|^2 \frac{\Delta(q-q'+q_b)}{\sinh \Big( \frac{ \hbar \omega_{q_bp_b}}{2 k_B T}\Big) }
 \times ( \delta(\omega_{q'p'}-\omega_{qp}+\omega_{q_bp_b})+\delta(\omega_{q'p'}-\omega_{qp}-\omega_{q_bp_b}))\\
&+\frac{\pi}{\hbar^2}\sum_{q_bp_b} \Big| F_{q q' q_b}^{p p' p_b}\Big|^2 \frac{\Delta(q+q'+q_b)}{\sinh\Big( \frac{ \hbar \omega_{q_bp_b}}{2 k_B T}\Big) }  \delta(\omega_{q'p'}+\omega_{qp}-\omega_{q_bp_b})+\delta_{qq'}\delta_{pp'}\frac{1}{\tau_{qp}}.
\end{align*}
$ F_{q q' q_b}^{p p' p_b}$ is the strength of the interaction in between the three phonons involved in the scattering \cite{chaput11} and $\Delta$ a function which is zero unless its argument is a reciprocal lattice vector, in which case it take the value $1$.\\
We can then make the following ansatz for $n_{qp}^{(1)}$, $\sinh\Big( \frac{ \hbar \omega_{qp}}{2 k_B T}\Big)n_{qp}^{(1)}\equiv f_{qp}\equiv\sum_{\alpha=1}^3 \int dt' \frac{\partial T(t')}{\partial r_{\alpha}} f_{qp}^{\alpha}(t-t') $,
where $\alpha$ is used to label the cartesian components of the vector $\vec{f}_{qp}(t-t')$. If equation \ref{eq:1} is transformed to Fourier space we obtain
\begin{align}
-i \omega \vec{f}_{qp}(\omega)+\frac{\hbar \omega_{qp}}{4 k_B T^2 \sinh\Big( \frac{ \hbar \omega_{qp}}{2 k_B T}\Big)} \vec{v}_{qp} = -\sum_{q'p'} \Omega'_{qp,q'p'} \vec{f}_{q'p'}(\omega). \label{eq:2}
\end{align} 
The velocity is odd under inversion, $\vec{v}_{-qp}=-\vec{v}_{qp}$, and it is easy to check that the collision matrix is even, $\Omega'_{-qp,-q'p'}=\Omega'_{qp,q'p'}$, which means that $\vec{f}_{-qp}(\omega)=-\vec{f}_{qp}(\omega)$. Because the Brillouin zone contains $q$ as well as $-q$ it shows that the collision matrix is not unique and that it is indeed possible to make other choices $\Omega_{qp,q'p'}$ such that $\sum_{q'p'}\Omega'_{qp,q'p'}\vec{f}_{qp}(\omega)=\sum_{q'p'}\Omega_{qp,q'p'}\vec{f}_{qp}(\omega) $ . We choose to work with $\Omega_{qp,q'p'}$ given by
\begin{align*}
\Omega_{qp,q'p'}&=\delta_{qq'}\delta_{pp'}\frac{1}{\tau_{qp}}+ \frac{\pi}{\hbar^2}\sum_{q_bp_b} \Big| F_{q q' q_b}^{p p' p_b} \Big|^2 \frac{\Delta(q+q'+q_b)}{\sinh \Big( \frac{ \hbar \omega_{q_bp_b}}{2 k_B T}\Big) } \\
&\times [ \delta(\omega_{q'p'}-\omega_{qp}+\omega_{q_bp_b})+\delta(\omega_{q'p'}-\omega_{qp}-\omega_{q_bp_b})+ \delta(\omega_{q'p'}+\omega_{qp}-\omega_{q_bp_b})]
\end{align*}
which can be obtained by a dummy change of variable $q' \to -q'$ in the summation of equation \ref{eq:2}. This matrix is clearly symmetric and can be shown to be positive definite using the same method than in\cite{liebfried}.\\
In the following we denote by $q$ a general point in the Brillouin zone, and by $k$ a point in the irreducible part of the Brillouin zone. $R$ are rotations of the isogonal point group $g$ of the crystal and $|g|$ denote the cardinal of that group. We denote by $g_k$ the multiplicity for the branches of the star of $k$.\\
In equation \ref{eq:2} if we restrict the velocity field to the irreducible part of the Brillouin zone the Boltzmann equation becomes
\begin{align*}
\frac{\hbar \omega_{kp}}{4 k_B T^2 \sinh\Big( \frac{ \hbar \omega_{kp}}{2 k_B T}\Big)} v_{kp}^{\alpha} = -\sum_{R' k' p'} (\Omega_{kp,R'k'p'}-i\omega \delta_{k,R'k'} \delta_{pp'})\frac{g_{k'}}{|g|} f_{R'k'p'}^{\alpha}(\omega).
\end{align*} 
Under the rotations $R$ the velocity transform like $ v_{Rkp}^{\alpha}=\sum_{\beta}R_{\alpha \beta}v_{kp}^{\beta}$ and it can be checked that the collision matrix is invariant, $\Omega_{Rkp,Rk''p'} =\Omega_{kp,k'p'} $. Therefore the Boltzmann equation written at point $Rk$ show that 
 $ f_{R'k'p'}^{\alpha}(\omega)$ and $ \sum_{\beta} R^{-1}_{\alpha \beta}f_{RR'k'p'}^{\beta}(\omega)$ fulfill the same equation. This gives $f_{Rkp}^{\alpha}(\omega)=\sum_{\beta}R_{\alpha \beta}f_{kp}^{\beta}(\omega)+u^{\alpha}(\omega)$, where $\vec{u}(\omega)$ is any vector in the null space of $\Omega_{kp,R'k'p'}-i\omega \delta_{k,R'k'} \delta_{pp'}$. We will see later that the null space of this operator does not contribute to the lattice thermal conductivity and therefore that we can choose $\vec{u}(\omega)=0$. In other words $\vec{f}_{kp}(\omega)$ transform like the velocity and can therefore be understood as being proportional to the phonon means free path. The Boltzmann equation can finally be reduced to an integral equation over the irreducible part of the Brillouin zone only, 
\begin{align*}
\frac{\hbar \omega_{kp}}{4 k_B T^2 \sinh\Big( \frac{ \hbar \omega_{kp}}{2 k_B T}\Big)} \sqrt{\frac{g_k}{|g|}} v_{kp}^{\alpha} = -\sum_{\beta k' p'} (\tilde{\Omega}_{\alpha kp,\beta k'p'}-i\omega \delta_{kk'} \delta_{pp'} P_{k'}^{\alpha \beta}) \sqrt{\frac{g_{k'}}{|g|}} f_{k'p'}^{\beta}(\omega)
\end{align*}
with
\begin{align*}
&\tilde{\Omega}_{\alpha kp,\beta k'p'}=\sqrt{\frac{g_k g_{k'}}{|g|}} \sum_{R'} R'_{\alpha \beta} \Omega_{ kp, R' k'p'} \,\,\,\,\,\,\, \text{and} \,\,\,\,\,\,\, P_{k}^{\alpha \beta}= \frac{g_k}{|g|} \sum_{R} R_{\alpha \beta} \delta_{k,Rk}.
\end{align*}
In matrix notation this is written as $| X \rangle = -(\tilde{\Omega}-i\omega  P) |f (\omega)\rangle$,
 with obvious definitions for $\tilde{\Omega}$ and $P$, and
 \begin{align*}
| X \rangle_{\alpha kp} =\frac{\hbar \omega_{kp}}{4 k_B T^2 \sinh\Big( \frac{ \hbar \omega_{kp}}{2 k_B T}\Big)} \sqrt{\frac{g_k}{|g|}} v_{kp}^{\alpha}   \,\,\,\,\,\,\, \text{and} \,\,\,\,\,\,\,
 |f (\omega)\rangle_{\alpha kp}= \sqrt{\frac{g_{k}}{|g|}} f_{kp}^{\alpha}(\omega).
\end{align*}
The operator $ P_{k}$ is working like the identity on vectors which transform like the velocity,  $P_{k} \vec{v}_{kp}=\vec{v}_{kp}$. $\tilde{\Omega}$ is a collision matrix. It tells that when working in the irreducible part of the Brillouin zone, and considering a transition from vector $k$ to $k'$, one should, obviously, also consider all the transitions to the different branches of the star of $k'$. Using the group properties of the set of rotation matrix, one can show that the matrices $P_{k}$ and $\tilde{\Omega}$ are symmetric.\\
The energy flux through the lattice is given by $\vec{J}_{E}(t)=\frac{1}{V} \sum_{qp} \hbar \omega_{qp} \vec{v}_{qp} n_{qp}(t)$ therefore its Fourier transform is $\vec{J}_{E}(\omega)=-\kappa(\omega) \frac{\partial T}{\partial \vec{r}} (\omega)$ with the thermal conductivity tensor given by
\begin{align*}
\kappa^{\alpha \beta}(\omega)=-\frac{1}{V}\sum_{qp} \frac{\hbar \omega_{qp}}{\sinh\Big( \frac{ \hbar \omega_{qp}}{2 k_B T}\Big)} v_{qp}^{\alpha} f_{qp}^{\beta}(\omega)=\frac{4 k_B T^2}{V}\sum_{qpq'p'}  f_{qp}^{\alpha}(\omega)(\Omega_{qp,q'p'}-i\omega \delta_{q,q'} \delta_{pp'})f_{q'p'}^{\beta}(\omega)
\end{align*}
In the second step we have used that the factor of $f_{qp}^{\beta}(\omega)$ in the summand is just the drift term in the Boltzmann equation.
As for the Boltzmann equation, the double integral over the Brillouin zone can be reduced to the irreducible part. We obtain
\begin{align*}
\kappa^{\alpha \beta}(\omega)& =\frac{4 k_B T^2}{V} \langle f(\omega) |  \mathcal{I}(\alpha,\beta)(\tilde{\Omega}-i \omega P) | f(\omega) \rangle
\end{align*}
with  $\mathcal{I}_{\gamma kp,\gamma' k' p'}(\alpha,\beta)=\delta_{kk'}\delta_{pp'} \sum_{R}R_{\alpha \gamma} R_{\beta \gamma'}$.
This operator is diagonal in the $kp$ space. Its value for the cartesian variables $\alpha$ and $\beta$ depend on the symmetry class of the system and can easily be calculated using the great orthogonality theorem of groups theory. From its definition it is also clear that $ \mathcal{I}(\alpha,\beta)= \mathcal{I}^t(\beta,\alpha)$ and that its purpose is to project out the components of the velocity and mean free path not involved it the $\alpha \beta$ component of the conductivity tensor.
The operator $\tilde{\Omega}(\alpha, \beta; \omega) \equiv \mathcal{I}(\alpha,\beta)(\tilde{\Omega}-i \omega P)$  appearing in the thermal conductivity  transform the same way as $\mathcal{I}$, $\tilde{\Omega}(\alpha, \beta; \omega)= \tilde{\Omega}^t(\beta, \alpha; \omega)$, therefore using the symmetry of $\tilde{\Omega}$ and $P$ we obtain the commutation relation $[\mathcal{I}(\alpha,\beta),\tilde{\Omega}-i \omega P]=0$.
This last identity implies the Onsager reciprocity relations at finite frequency, $\kappa^{\alpha \beta}(\omega)=\kappa^{\beta \alpha}(\omega)$
and therefore allow to obtain a more symmetric equation for the thermal conductivity, 
\begin{align*}
\kappa^{\alpha \beta}(\omega)=\frac{2 k_B T^2}{V} \langle f(\omega) |(\mathcal{I}(\alpha,\beta)+ \mathcal{I}(\beta, \alpha))(\tilde{\Omega}-i \omega P) | f(\omega) \rangle,
\end{align*}
because $\mathcal{I}(\alpha,\beta)+ \mathcal{I}(\beta, \alpha)$ is a symmetric matrix. It shows also that the null space of $(\mathcal{I}(\alpha,\beta)+ \mathcal{I}(\beta, \alpha))(\tilde{\Omega}-i \omega P)$ does not contribute to the lattice thermal conductivity. A vector which belongs to the null space of $\tilde{\Omega}-i \omega P$ also belongs the null space of  $(\mathcal{I}(\alpha,\beta)+ \mathcal{I}(\beta, \alpha))(\tilde{\Omega}-i \omega P)$. Therefore we can choose a solution to the Boltzmann equation which is orthogonal to ker$(\tilde{\Omega}-i \omega P)$ and transform like the velocities, $|f (\omega)\rangle =-(\tilde{\Omega}-i\omega  P)^{\sim 1}| X \rangle$.
Here $\sim 1$ is used to denote the Moore-Penrose inverse. The thermal conductivity can now be expressed as an average value over the known vector $| X \rangle$, 
\begin{align*}
\kappa^{\alpha \beta}(\omega)=\frac{2 k_B T^2}{V} \langle X |(\tilde{\Omega}-i\omega  P)^{\sim 1}(\mathcal{I}(\alpha,\beta)+ \mathcal{I}(\beta, \alpha)) | X \rangle.
\end{align*}
The matrices $\mathcal{I}(\alpha,\beta)+ \mathcal{I}(\beta, \alpha)$ and $\tilde{\Omega}$ are symmetric and commutes. It is therefore possible to find a set of eigenvectors $| e_r \rangle$ such that $\tilde{\Omega} | e_r \rangle = \omega_r  | e_r \rangle$  and $(\mathcal{I}(\alpha,\beta)+ \mathcal{I}(\beta, \alpha)) | e_r \rangle = i_r(\alpha,\beta)  | e_r \rangle$.
This gives finally a spectral representation for the dynamical thermal conductivity, 
\begin{align*}
\kappa^{\alpha \beta}(\omega)=\frac{2 k_B T^2}{V} \sum_{r}' \frac{i_r(\alpha,\beta)| \langle X | e_r \rangle |^2}{\omega_r -i \omega}=\int d \omega' \frac{\rho_{\alpha \beta}(\omega')}{\omega'-i \omega}
\end{align*}
where $\rho_{\alpha \beta}(\omega')$ is a spectral density and  the prime in the summation tells that the null space have to be excluded.\\
The method has been applied to materials with high (diamond), medium (silicon) and low (magnesium silicide) thermal conductivity. \emph{Ab initio} calculations are performed to obtain the interaction strength in between the phonons, $ F_{q q' q_b}^{p p' p_b}$, using the method in \cite{chaput11}. Therefore no adjustable parameters are used in the calculation. A $15 \times 15 \times 15 \times$ mesh is used to sample the Brillouin zone and the scattering by isotopes and surfaces have been included in the usual ways \cite{ward09}. The results of the calculations are shown on figure \ref{fig:1}a for the static thermal conductivity.  In each case the agreement with experiment is excellent. This has to be related to the use of the previous equation which only involves the diagonalization of small matrices and where symmetry has been used at best to reduce numerical uncertainties.\\
The real $(\kappa_r)$ and imaginary parts $(\kappa_i)$ of the dynamical lattice thermal conductivity at room temperature are show on figure \ref{fig:1}b as a function of frequency. A rapid drop of $\kappa_r$ is observed after some cut off frequency $1/\tau_0$ which also correspond to a maximum in $\kappa_i$. Such rapid decrease for $\kappa_r$ has already been obtained in silicon using molecular dynamics calculations\cite{volz01}, and is known for long time \cite{chester}, in principle. When considering rapid time variations, the Fourier's law has to be modified to account for the finite time needed to establish a current, what may eventually lead to thermal waves at large frequency. More work is needed to obtain the phonon second sound of \cite{sham67}, but fortunately the structure of the equations does not change and our procedure can be applied. Such work is in progress.\\
The evolution of $1/\tau_0$ with temperature is shown in the inset of figure \ref{fig:1}a. It is clear that at high enough temperature it becomes proportional to the temperature, as required for a relaxation time. It can also be seen on figure 1b that the spread of $\kappa_i$ looks broader for silicon. This may indicate that a single relaxation time is not sufficient to account for the dynamics and would leads to a more complicated evolution of the temperature. This is indeed confirm by the spectral density $\rho_{\alpha \beta}$ ploted in the inset of figure 1b since there is a significant weight around $5 \, 10^8$Hz. \\
To summarize, exploiting the symmetry of the system we have given a solution to the Boltzmann equation which allow to compute the thermal conductivity from a spectral representation. This way numerical errors are greatly reduced and therefore the study of material with low thermal conductivity is possible. In addition it is no more difficult to obtain the dynamical thermal conductivity which is calculated here for the first time. This allow for a quantitative estimate of the drop frequency of $\kappa_r$. This can be of great interest in thermoelectric applications where $\kappa$ need to be reduced to increase ZT, but also for the industry of microprocessors. Considering their clock rate, heat transport at high frequency need to be understood. The drop frequency of the materials calculated here are very large. However a decrease of $\kappa_r$ at much lower frequencies has been reported experimentally in alloys compounds\cite{koh07} and could lead to applications. 
\bibliography{boltzmann}
\end{document}